\documentclass[aps,prb,twocolumn,amsmath,amssymb,nofootinbib,superscriptaddress,floatfix]{revtex4}

\usepackage{amsmath}
\usepackage{amssymb}
\usepackage{amsthm}
\usepackage[pdftex]{color}
\usepackage{bm} 
\usepackage{ulem}   
\newcommand{\bs}[1]{{\boldsymbol{#1}}}

\begin{document}

\title{ 
Erratum for ``Elementary formula for the Hall conductivity of interacting systems''
      }

\date{\today}

\author{Titus Neupert} 
\affiliation{
Condensed Matter Theory Group, 
Paul Scherrer Institute, CH-5232 Villigen PSI,
Switzerland
            } 

\author{Luiz Santos} 
\affiliation{
Department of Physics, 
Harvard University, 
17 Oxford St., 
Cambridge, MA 02138,
USA
            } 

\author{Claudio Chamon} 
\affiliation{
Physics Department, 
Boston University, 
Boston, Massachusetts 02215, USA
            } 

\author{Christopher Mudry} 
\affiliation{
Condensed Matter Theory Group, 
Paul Scherrer Institute, CH-5232 Villigen PSI,
Switzerland
            } 

\begin{abstract}
When deriving a formula for the Hall conductivity of interacting electrons 
in Ref.~\onlinecite{us}, we have relied on an unjustified implicit assumption that 
a certain gauge choice could be made. Only under this condition would 
the formula follow. If this condition fails, the formula we derived 
does not lead to the exactly quantized value of the Hall conductance 
in fractional Chern insulators. 
\end{abstract}

\maketitle

As pointed out in Ref.~\onlinecite{fdp}
by Simon, Harper, and Read, 
and independently by Haldane~\cite{Haldane}
the formula~(4.15) for the Hall resistivity 
in Ref.~\onlinecite{us} is not invariant under certain symmetries.
In the case of Ref.~\onlinecite{fdp}, the symmetry refers to the arbitrariness in defining the phase of the Blochs states in each band, given an orbital basis.
Haldane refers to the choice of how to embed the orbitals of a tight-binding Hamiltonian in position space, given a Bloch basis.
Under these transformations, the Hamiltonian is invariant, but the Hall conductivity, when computed with formula~(4.15) from Ref.~\onlinecite{us}, would be variant. Both of these transformations are gauge transformations on the basis of the full single-particle Hilbert space. 
Here, we trace back the loss of gauge invariance
as the result of a choice of gauge,
whose existence was implicitly assumed but not proven.

In Sec.~IV of Ref.~\onlinecite{us}, 
we make the decomposition
[Eq.~(4.8)] of the position operator
\begin{equation}
\bs{X}=
\bs{T}
+ 
\bs{A}.
\label{eq: decomposition X}
\end{equation} 
As we remarked in the paper, the decomposition%
~(\ref{eq: decomposition X}) is not unique, but basis dependent.
Under basis transformations of the single-particle Hilbert space, the
operator $\bs{A}$ transforms like an operator-valued gauge field. The
position operator $\bs{X}$ is invariant under such gauge
transformations.

In establishing the formula for the conductance, we used that $\bs{T}$
generates translations in momentum space, and moves the ground states
away from a superposition of Fock states with definite total
momentum $\bs{Q}^{\,}_{0}$. In contrast, the operator $\bs{A}$ does not
shift momentum, for it is diagonal in the single-particle momentum
$\bs{k}$. In other words, we \textit{assumed}
 $\bs{T}$ is \textit{strictly} off-diagonal in $\bs{k}$-space, while $\bs{A}$ is
diagonal in $\bs{k}$-space.

All formal manipulations in Sec.~IV follow from this 
additive decomposition into two
operators,
one that shifts and the other that does not
not shift the total momentum quantum number (assumed to be a good
quantum number) of the exact ground state. Notice that if such
decomposition is possible, it is not invariant under gauge
transformations. A gauge transformation will generically add a
diagonal in $\bs{k}$ contribution to $\bs{T}$ that is compensated by
an opposite shift in $\bs{A}$. We have thus fixed a gauge by
requiring that $\bs{T}$ is {\it strictly} off-diagonal in
$\bs{k}$-space. Therefore, the expression that we obtain for the
conductance is not gauge invariant, but gauge fixed.

Is this choice of gauge possible for the exact ground state
of any Hamiltonian satisfying the assumptions made in
Ref.~\onlinecite{us}? 
We have not shown it is in
Ref.~\onlinecite{us}. This is therefore an unjustified step. 
If it is not possible to select such a gauge that permits
the manipulations in Sec. IV that involved translations in momentum
space and their effects on projected subspaces of definite momentum
such as the subspace of ground states, then the formula is not
valid. 

The manipulations of Sec.~IV allowed us to dispose of the
contributions from the $\bs{T}$ operator defined by
Eq.~(\ref{eq: decomposition X}). If 
there are contributions from $\bs{T}$ to the Hall conductivity
(because the gauge condition required to ignore $\bs{T}$ is not 
permissible), then the formula we derived will never be exact. 
It could at best serve as an approximation to the quantized Hall conductance.
We remark that, if approximate, the expression is close to the
  quantized value in all systems that we have thus far studied
  numerically.

\section*{Acknowledgments}

We thank Simon, Harper, and Read for sharing their criticisms on the
formula we had derived. It was because of their arguments that we
revisited our derivation of a formula for the Hall conductivity
and realized that we had assumed a gauge-fixing condition 
without proving it.
They took a step further in their effort in
Ref.~\onlinecite{fdp} to analyze the consequences to 
the quantization of our formula for the Hall conductivity
at which we had arrived.
Furthermore, we are grateful to F.D.M.\ Haldane for useful discussions.


\begin{thebibliography}{99}

\bibitem{us}
T.\ Neupert, L.\ Santos, C.\ Chamon, and C.\ Mudry, 
Phys. Rev. B \textbf{86}, 165133 (2012).

\bibitem{fdp}
S.\ Simon, F.\ Harper, and N.\ Read,
submitted comment.

\bibitem{Haldane}
F.D.M.\ Haldane,
private communication.

\end{thebibliography}
\end{document}